\documentclass[aps,pra,amsfonts,twocolumn,superscriptaddress]{revtex4-1}
\usepackage{epsfig}
\begin{document}
\title{Quadratic Entanglement Criteria for Qutrits}
\author{Krzysztof Roso\l ek}\affiliation{Institute of Theoretical Physics and Astrophysics, Faculty of Mathematics, Physics, and Informatics, University of Gda\'nsk, ul. Wita Swosza 57, 80-308 Gda\'nsk, Poland}
\author{Marcin Wie\'sniak}\affiliation{Institute of Informatics, Faculty of Mathematics, Physics, and Informatics, University of Gda\'nsk,\\ ul. Wita Swosza 57, 80-952 Gda\'nsk, Poland}
\author{Lukas~Knips}\affiliation{Max-Planck-Institut f\"ur Quantenoptik, Hans-Kopfermann-Strasse 1, D-85748 Garching, G, Germany}\affiliation{Department f\"ur Physik, Ludwig-Maximilians-Universit\"at, D-80797 M\"unchen, Germany}
\begin{abstract}	
The problem of detecting non-classical correlations of states of many qudits is incomparably more involved than in a case of qubits. The reason is that for qubits we have a convenient description of the system by the means of the well-studied correlation tensor. Simply, the complete information about the state can be encoded in mean values of dichotomic measurements. We demonstrate that for three-dimensional quantum subsystems we are able to formulate nonlinear entanglement criteria of the state with existing formalisms. We also point out where the idea for constructing these criteria fails for higher-dimensional systems, which poses well-defined open questions.
\end{abstract}
\maketitle
\section{Introduction}
Quantum correlations are known to be capable of outperforming classical ones. While separable states can be perfectly correlated in one way at a 	time, entangled ones may reveal perfect correlations, say, whenever the same quantity is measured by two observers. This observation has lead to a serious debate about the most fundamental aspects of nature. First, Einstein, Podolsky, and Rosen \cite{EPR} have asked if quantum mechanics can be supplemented with additional, hidden parameters, and later it was answered that if it was indeed so, these parameters would need to go beyond certain reasonable requirements, such as locality \cite{Bell}, or noncontextuality \cite{KochenSpecker}. 

The Bell theorem \cite{Bell} has consequences of not only philosophical nature, but has also found applications in certain communication tasks. In particular, having a Bell inequality violated by a quantum state is equivalent to an advantage in a distributed computing \cite{Communication}. Specifically, if protocol users share an entangled state, they can achieve a higher probability of locally getting the correct value of a certain function than when they are allowed only to communicate classically. The role of the Bell theorem has been also pointed out in the context of, e.g., cryptography \cite{Ekert}.

Therefore, schemes of entanglement detection have gathered a lot of attention for both fundamental and practical reasons. The task is very simple for pure states, which practically never occur in a real life. However, for mixed states it is still an open question. One method is to apply a positive, but not a completely positive map to one of subsystems \cite{Peres,Horod}. This should drive an entangled state out of the set of physically admissible density operators. By the Jamio\l kowski-Choi isomorphism \cite{JC}, we can equivalently use an entanglement witness, a composite observable taking negative mean values only for entangled states. In this manner, we can certify all forms of entanglement, but we do not know all the non-completely positive maps. In order to make entanglement detection schemes more efficient, nonlinear criteria were introduced. They appeared also in particular context of necessary conditions on states to violate Bell inequalities \cite{WW,WZ,ZB}. 
A state can violate Werner-Wolf-Weinfurter-\.Zukowski-Brukner inequalities only if (but not necessarily if) certain of its squared elements of the correlation tensor add up to more than 1. A similar condition appeared in the context of so-called geometrical inequalities \cite{Nagata}, which treat correlations of the system as multidimensional vector not belonging to a convex set of local realistic models. This approach resulted in geometrical entanglement criteria \cite{WL}, which are highly versatile, and quadratic ones, particularly easy to construct \cite{WIESNMARU,Lukas,WLask}.

Up to date, these methods turn out to be successful mainly for collections of qubits, as their states are conveniently described by the means of the correlation tensor. The deficit of Bell inequalities and entanglement criteria for higher-dimensional constituents of quantum systems follow also from our inability to generalize this tool. Pauli matrices, the foundation of this achievement, have many interesting properties, each contributing to the success. They are Hermitian, unitary, traceless, for individual subsystems their measurements are complete (except for the unit matrix), meaning that the individual mean values contain the full information about the statistics of outcomes, and they have unbiased bases as their eigenbases. In contrast, one of the straight-forward generalizations, the Gell-Mann matrices, do not satisfy any commutativity relations. If we used them to create a correlation tensor, its elements would not be independent from one another.

In this contribution we show that the notions known for the formalism of the tensor product for multiqubit states can be straight-forwardly stretched to qutrits, when we associate complex root of infinity to local measurement outcomes. In particular, this generalized tensor product is a subject to linear and quadratic bounds. Basing on these bounds, we can derive quadratic (and geometrical) entanglement criteria. For higher-dimensional systems, this is still an open challenge.
\section{Formalism of Many-qubit states}
As we have already mentioned, the success of describing and analyzing the states of many qubits is due to the particularly convenient representation through a correlation tensor. Its  elements are mean values of tensor products of Pauli matrices, $T_{\bar{i}}=\langle o_{\bar{i}}\rangle$, $o_{\bar{i}}=\sigma_{i_1}^{[1]}\otimes\sigma_{i_2}^{[2]}\otimes...$, $\bar{i}={i_1,i_2,...,i_{N}}$, and
\begin{eqnarray}
\sigma_0=\left(\begin{array}{cc} 1&0\\ 0&1\end{array}\right),\,&&
\sigma_1=\left(\begin{array}{cc} 0&1\\ 1&0\end{array}\right),\nonumber\\
\sigma_2=\left(\begin{array}{cc} 0&-i\\ i&0\end{array}\right),\,&&
\sigma_3=\left(\begin{array}{cc} 1&0\\ 0&-1\end{array}\right),\\
\end{eqnarray}
Operators $o_{\bar{i}}$ form an orthonormal basis, $(o_{\bar{i}},o_{\bar{j}})=\text{tr} o_{\bar{i}}o_{\bar{j}}=2^N\delta_{\bar{i},\bar{j}}$. This orthogonality can have three different reasons. When either $o_{\bar{i}}$ or $o_{\bar{j}}$ is the unit matrix, the other operator is traceless. When $o_{\bar{i}}$ and $o_{\bar{j}}$ commute, but differ from each other and neither of them is the unit matrix, their eigenvalues are distributed in such a way that their product adds up to zero. Finally, when they do not commute, they anticommute and their eigenbases can be chosen to be unbiased, i.e., the scalar product between any vector from one basis and any one from the other is constant in modulo. 
For a given state $\rho$, let the correlation tensor be a set of averages $\{T_{\bar{i}}\}=\{\text{tr}\rho o_{\bar{i}}\}$. Naturally, $T_{00...0}\equiv 1$, but also for a single qubit we have the pronounced complementarity 	relation \cite{complementarity},
\begin{equation}
\sum_{i=1}^{3}\langle \sigma_i\rangle^2\leq 1.
\end{equation}
This relation can be straight-forwardly generalized to any set of mutually anticommuting operators (where $Z$ is some set of multiindex values),
\begin{eqnarray}
&\{o_{\bar{i}},o_{\bar{j}}\}_{\bar{i},\bar{j}\in Z}\propto \delta_{\bar{i},\bar{j}}\Rightarrow\nonumber\\
&\sum_{\bar{i}\in Z}T^2_{\bar{i}}\leq 1.
\end{eqnarray}
Notice that operators $o_{\bar{i}}$ and $o_{\bar{j}}$  anticommute iff superindices differ on odd number of positions, excluding those, where one superindex has ``0''. In Ref. \cite{WIESNMARU} this property was further generalized to cut-anticommutativity. Namely, consider two operators, $o_1=o_1^{[A]}\otimes o_1^{[B]}$ and $o_2=o_2^{[A]}\otimes o_2^{[B]}$. We say that they anticommute with respect to cut $A|B$ if they anticommute on either of the subsystem. Consequently, 
\begin{eqnarray}
\label{comp1}
&\{o_1,o_2\}_{A|B}=0\Rightarrow\nonumber\\
&\langle o_1\rangle^2+\langle o_2\rangle^2\leq 1
\end{eqnarray}
for states, which are factorizable (and, by convexity, thus separable) with respect to the cut. This lead in Ref. \cite{WIESNMARU} to constructing quadratic entanglement criteria based solely on anticommutativity properties of operators. The main goal of this contribution is to show that the formalism for qutrits can also be used for this purpose. 
\section{Correlation Tensor Formalism for Many Qutrits}
We are now looking for a description of a qutrit, in which each measurement gives us a complete information about the probability distribution of three  outcomes. To remove any dependencies, we expect the measurements on individual qutrits used for establishing the correlation tensor to be have mutually unbiased bases (MUBs) as their eigenbases. Lastly, since we want to formulate the complementarity relation similar to Eq. (\ref{comp1}), so we expect the eigenvalues to be of modulo 1. 
A family satisfying these requirements for three-dimensional subsystems are the Heisenberg-Weyl matrices. They are given as
\begin{eqnarray}	 
h_0=\left(\begin{array}{ccc} 1&0&0\\0&1&0\\0&0&1\end{array}\right),\nonumber\\
h_1=\left(\begin{array}{ccc} 1&0&0\\0&\omega&0\\0&0&\omega^2\end{array}\right),\, && h_2=\left(\begin{array}{ccc} 0&1&0\\0&0&1\\1&0&0\end{array}\right),\nonumber\\
h_3=\left(\begin{array}{ccc} 0&1&0\\0&0&\omega\\\omega^2&0&0\end{array}\right),\, && h_4=\left(\begin{array}{ccc} 0&1&0\\0&0&\omega^2\\\omega&0&0\end{array}\right),\nonumber\\
h_5=h_1^\dagger,\,&&h_6=h_2^\dagger,\nonumber\\
h_7=h_3^\dagger,\,&&h_8=h_4^\dagger
\end{eqnarray}	
($\omega=\exp(2\pi i/3)$).

First, let us show that this representation of a state is complete, that is, the data can be used for state tomography. As given in Ref. \cite{MUBS1}, a state can be given as
\begin{equation}
\label{Tomo1}
\rho=-1+\sum_{m=1}^4\sum_{k=0}^{2}p(m,k)|mk\rangle\langle mk|,
\end{equation}
where $m$ enumerates the mutually unbiased basis, the eigenbasis of $h_m$, $|mk\rangle$ is the $k$th state of this basis and $p(m,k)=\langle mk|\rho|mk\rangle$. Now, consider the following quantity:
\begin{equation}
T_m=\text{tr}\rho h_m^\dagger.
\end{equation}
For simplicity, let us represent complex numbers and operators as vectors, i.e., $\vec{a}=(\text{Re}a,\text{Im}a)$ and $\vec{o}=1/2(o+o^\dagger, -i(o-o^\dagger))$. Furthermore, let us denote 1,$\omega, \omega^2$ as $\vec{v}_0=(1,0), \vec{v}_1=\left(-\frac{1}{2},\frac{\sqrt{3}}{2}\right), \vec{v}_2=\left(-\frac{1}{2},-\frac{\sqrt{3}}{2}\right)$. Notice that this defines a new scalar product, which leads to
\begin{eqnarray}
&& \vec{T}_m\cdot\vec{o}_m\nonumber\\
=&&\left(\sum_{k=0}^{2} p(m,k)\vec{v}_k\right)\cdot\left(\sum_{l=0}^{2}\vec{v}_l|ml\rangle\langle ml|\right)\nonumber\\
=&&-\frac{1}{2}+\frac{3}{2}p(m,k)|mk\rangle\langle mk|,\nonumber\\
&&\sum_{k=0}^{2} p(m,k)|mk\rangle\langle mk|\nonumber\\
=&&\frac{2\vec{T}_m\vec{o}_m+1}{3},\nonumber\\
\rho=&&\sum_m\frac{2}{3}\vec{T}_m\vec{o}_m-\frac{1}{3}.
\end{eqnarray}
The last equation can be plugged in to Eq. (\ref{Tomo1}). When the usual tensor product is used, this formula is extended by replacing products of probabilities with joint probabilities, $p(k,m)p(l,n)\rightarrow p(k,m,l,n)=\langle k,m|\otimes\langle l,n|\rho|k,m\rangle\otimes|l,n\rangle$. 

Let us now consider the complementarity relations between tensor products of Heisenberg-Weyl operators. For certain noncommuting groups of operators, $\{o_{\bar{j}}\}_{\bar{j}}$, we shall have
\begin{equation}
\label{comple}
\sum_j|\langle o_{\bar{j}} \rangle|^2\leq 1,
 \end{equation}
the equivalent of which was one of the key ingredients of Ref. \cite{WIESNMARU} for qubits. Therein, this complementarity directly follows from the anticommutativity relations between the various Pauli matrix tensor products. Here, the situation is not as simple. The argument cannot go through directly as Heisenberg-Weyl tensor product operators do not anticommute. Still, we find some forms of complementarity between these operators. For an individual qutrit we shall have
\begin{eqnarray}
\label{proof1}
1\leq&&\text{tr}\rho^2\nonumber\\
=&&\sum_{i,j=1}^3|\rho_{ij}|^2\nonumber\\
=&&\frac{1}{9}\sum_{i,j=0}^2\left|\left\langle\sum_{k=0}^2\omega^ih_2^jh_1^{jk}\right\rangle\right|^2\nonumber\\
=&&\frac{1}{3}\sum_{i,j=0}^2|\langle h_1^ih_2^j\rangle|^2,\nonumber\\
3\leq&&1+2\sum_{(i,j)\in\{(1,0),(0,1),(1,1),(1,2)\}}|\langle h_1^ih_2^j\rangle|^2,
\end{eqnarray} 
where the transition between the third and the fourth line comes from the Parseval's theorem for the Fourier transform.

Now, we are ready to consider the complementarity for many-qutrit operators. Here our possibilities are quite limited. One would expect that as long as tensor products do not commute, the sum of squared moduli of their averages for any state would not exceed 1. This is false, however. We have found 792 distinguished sets of seven mutually non-commuting two-qutrit operators, $\{o_i\}_{i=1}^7$, and found that for all of them there exist states, for which $\sum_{i=1}^7|\langle o_i\rangle|^2=\frac{5}{4}$. For the complete set of two-tensor products of Heisenberg-Weyl operators, from the semi-positivity of the state one can show that
\begin{equation}
\sum_{i,j=0}^8|\langle h_i\otimes h_j\rangle|^2\leq 9.
\end{equation}
Nevertheless, we can easily argue for the complementarity of a smaller set. In particular consider a pair of operators, $o_1$ and $o_2$, which do not commute with each other. By diagonalization one of them and a proper choice of phases of the new basis states we can bring to the $3\times 3$ block-diagonal form, where each of the blocks takes form
\begin{eqnarray}
\label{proof2}
&[o_1]_{\text{block}}= h_1,&\nonumber\\
&[o_2]_{\text{block}}\propto h_2&
\end{eqnarray}
and the complementarity follows directly from Eqs. (\ref{proof1}). In addition, one may have two more operators, the blocks of which correspond to $h_3$ $(h_7)$ and $h_4$ $(h_8)$, up to global phases, extending the complementarity principle from two general to four specific operators. Notice that the operation diagonalizing $o_1$ does not need to be local, so this complementarity is not of a strictly local nature.

We can now transplant the rest of ingridients from Ref. \cite{WIESNMARU} to this consideration. Obviously, if we have mutually commuting operators, it suffices to choose a common eigenstate of all of them, to have all the mean values equal to 1. Also, we can use the proof from the reference that in quadratic entanglement criteria, mixing states cannot improve the situation.

Another fact we need for the construction is that for product states $\rho=\rho_1^{[A]}\otimes\rho_2^{[B]}$ and a multiqutrit operator in form $\vec{O}=o_1^{[A]}\otimes o_2^{[B]}$, where $[A]$ and $[B]$ are subsystems, we have
\begin{equation}
|\langle \vec{O}\rangle|^2=|\langle \vec{o}_1^{[A]}\rangle|^2|\langle \vec{o}_2^{[B]}\rangle|^2,
\end{equation}
which, again follows directly from the correspondence between the two-dimensional vector eigenvalues and the complex root-of-unity eigenvectors. However, this relation fails for $d>3$, when we replace the complex roots of unity as eigenvalues with $(d-1)$-vectors $\vec{v}_{d,i}$ satisfying relation
\begin{equation}
\vec{v}_{d,i}\cdot\vec{v}_{d,j}=\frac{d\delta_{i,j}-1}{d-1}.
\end{equation}
Thus our method is applicable only for a collection of qutrits.

\section{Examples}

Consider the four-qutrit GHZ state, which in the computational basis ($h_1|i\rangle=\omega^i|i\rangle$) has form
\begin{equation}
|\text{GHZ}_{3,4}\rangle=\frac{1}{\sqrt{3}}\sum_{j=0}^2|iiii\rangle.
\end{equation} 
Prefect correlations of this state include (hereafter, we omit the tensor 	product signs)
\begin{eqnarray}
\label{GHZCorrs}
&\langle h_2h_2h_2h_2\rangle=\langle \Pi(h_1h_5h_1h_5)\rangle&=1,
\end{eqnarray}
where $\Pi(abcd)$ denotes an arbitrary permutation of $a,b,c,d$ in terms of the tensor product. Hence we can use the criterion
\begin{equation}
|\langle h_1h_5h_1h_5\rangle|^2+|\langle h_2h_2h_2h_2\rangle|^2\leq_{\text{SEP}}1\leq 2 (\text{for }|GHZ_{3,4}\rangle)
\end{equation}
to exclude separability with respect to bipartitions $AB|CD$ and $AD|BC$, while the criterion 
\begin{equation}
|\langle h_1h_1h_5h_5\rangle|^2+|\langle h_1h_5h_1h_5\rangle|^2\leq_{\text{SEP}}1\leq 2 (\text{for }|GHZ_{3,4}\rangle)
\end{equation}
can used to exclude separability between subsystems $AC$ and $BD$. Additionally, both of these criteria are sensitive to all one-versus-three cuts. Thus, a simultaneous violation of both of these inequalities certifies true multipartite entanglement of the tested state (in principle, different from the GHZ state).

The next example is the four-qutrit cluster state,
\begin{eqnarray}
|C_{3,4}\rangle=&&\frac{1}{3}\sum_{i,j=0}^2\omega^{ij}|ijij\rangle,
\end{eqnarray}
for which we can utilize correlations
\begin{eqnarray}
&\langle h_0h_2h_5h_2\rangle=\langle h_2h_0h_2h_5\rangle&\nonumber\\
=&\langle h_5h_2h_0h_2\rangle=\langle h_2h_5h_2h_0\rangle&=1,
\end{eqnarray}
Which gives us the following criterion for true four-qutrit entanglement:
\begin{eqnarray}
&\frac{1}{2}\left(|\langle h_0h_2h_5h_2\rangle|^2+|\langle h_2h_0h_2h_5\rangle|^2\right)+\nonumber\\
&\frac{1}{2}\left(|\langle h_5h_2h_0h_2\rangle|^2+|\langle h_2h_5h_2h_0\rangle|^2\right)&>1.
\end{eqnarray}
Again, these four correlations can be, in principle established together, and while measuring in local MUBs, it again takes only two series of measurements to establish all four of them.

\begin{figure}
	\centering
		\includegraphics{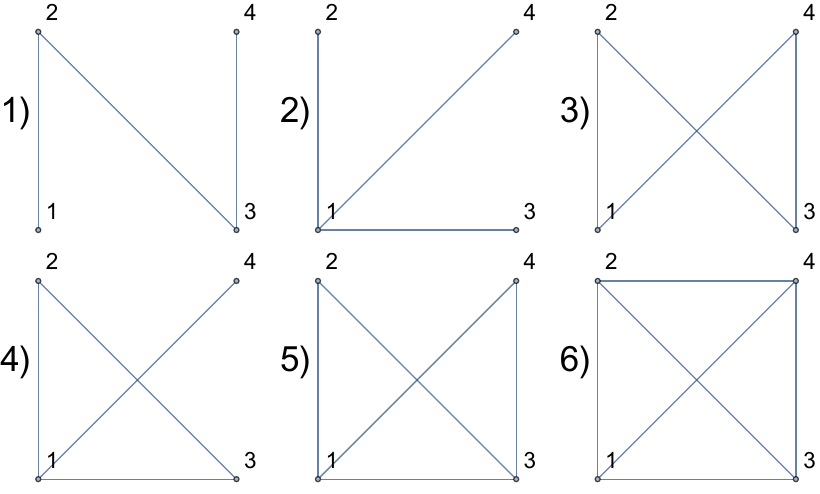}
	\caption{Layouts of all four qutrit graph states. Qutrits are represented by vertices, while edges symbolize the application of the generalized control-$Z$ operation of Eq. (\ref{eq:controlZ}).}
	\label{fig:Graphs}
\end{figure}

To demonstrate the usefulness and convenience of our method, let us consider four-qutrit graph states in general. Imagine a collection of four qutrits, each initialized in state $\frac{1}{\sqrt{3}}(|0\rangle+|1\rangle+|2\rangle)$. Now we take a graph, which connects four vertices. There are two such graphs with three edges (a path and a three-arm star), two with four (a loop and a triangle with a leg), one with five (a loop with one diagonal) and the complete graph has six edges. The graphs are presented in Fig. \ref{fig:Graphs}. If two qutrits are connected by an edge on the graph, we entangle them by applying a generalization of the control-$Z$ operation,

\begin{equation}
Ch_1=\text{Diag}(1,1,1,1,\omega,\omega^2,1,\omega^2,\omega).
\label{eq:controlZ}
\end{equation}
Each four-qutrit graph state has a total of 80 perfect correlations for nontrivial tensor products of the $h$ operators. Lists of these correlations have been made, and from them we choose triples of operators, which satisfy the following conditions: (i) their mean value for the reference state has the absolute value equal to 1, (ii) for every bipartite cut, at least one pair does not cut-commute, (iii) they can be established in two measurement series. We came to the conclusion that the true four-partite entanglement is certified if
\begin{eqnarray}
&&\text{for Graph 1:}\nonumber\\
&&|\langle h_3h_8h_4h_7\rangle|^2+|\langle h_6h_0h_2h_5\rangle|^2>1\nonumber\\
\wedge&&|\langle h_3h_8h_4h_7\rangle|^2+|\langle h_0h_5h_2h_5\rangle|^2>1,\nonumber\\
&&\text{for Graph 2:}\nonumber\\
&&|\langle h_2h_5h_5h_5\rangle|^2+|\langle h_1h_6h_6h_4\rangle|^2>1\nonumber\\
\wedge&&|\langle h_2h_5h_5h_5\rangle|^2+|\langle h_5h_6h_6h_0\rangle|^2>1,\nonumber\\
&&\text{for Graph 3:}\nonumber\\
&&|\langle h_3h_3h_3h_3\rangle|^2+|\langle h_1h_2h_1h_2\rangle|^2>1\nonumber\\
\wedge&&|\langle h_3h_3h_3h_3\rangle|^2+|\langle h_1h_0h_1h_6\rangle|^2>1,\nonumber\\
&&\text{for Graph 4:}\nonumber\\
&&|\langle h_2h_5h_5h_5\rangle|^2+|\langle h_4h_3h_3h_7\rangle|^2>1\nonumber\\
\wedge&&|\langle h_2h_5h_5h_5\rangle|^2+|\langle h_8h_3h_0h_3\rangle|^2>1,\nonumber\\
&&\text{for Graph 5:}\nonumber\\
&&|\langle h_4h_2h_6h_2\rangle|^2+|\langle h_0h_3h_7h_1\rangle|^2>1\nonumber\\
\wedge&&|\langle h_4h_2h_6h_2\rangle|^2+|\langle h_3h_7h_0h_5\rangle|^2>1,\nonumber\\
&&\text{for Graph 6:}\nonumber\\
&&|\langle h_2h_8h_8h_8\rangle|^2+|\langle h_0h_3h_3h_3\rangle|^2>1\nonumber\\
\wedge&&|\langle h_2h_8h_8h_8\rangle|^2+|\langle h_3h_3h_3h_0\rangle|^2>1.\nonumber\\
\end{eqnarray}
Notice that not all of these correlations are equal to 1, but since the criteria are quadratic, this is no concern.
\section{Conclusions}
We have shown a generalization of a derivation of graph-based quadratic entanglement criteria, known already for qubits, to qutrits. It was achieved by using Heisenberg-Weyl operators, which can be seen vector-valued observables. While the obtained criteria can be applied to a relatively small set of states, namely those with very strong correlations, they are easy to derive, compared to most other methods. One does not need to optimize over the whole set of product states, but simply find some pairs of correlations, that we expect to be simultaneously high. This was well demonstrated in case of four-qutrit graph states. 

There are some differences between the derivation presented in Ref. \cite{WIESNMARU} and the above. Therein, we enjoyed the complementarity relation for an arbitrarily large set of cut-anticomming operators. For qutrits, we have found counterexamples. The complementarity principle holds in general for pairs of (cut-)noncommuting observables, and for more only in special cases. One still can, however, construct criteria such as those in Ref. \cite{Lukas}, involving only two terms each. For a given term, we take as many pairs as necessary to exclude separability of the state along all cuts.

Interestingly, we were not able to push the reasoning even further, to dimensions of subsystems higher than 3. There are few obstacles in generalizing the proofs. One is that the proof of the complementarity relation (Eq. (\ref{comple})) explicitly refers to the Heisenberg-Weyl formalism, which consists of the shift operators. The other difficulty is that for vector eigenvalues, we were unable to derive a dependence between the length of the mean value of a joint observable and the lengths for local operators.
\section{Acknowledgements}
The work is subsidized form funds for science for years 2012-2015 
approved for
international co-financed project BRISQ2 by Polish Ministry of Science 
and Higher Education (MNiSW). This work is a part of BRISQ2 Project co-financed by EU. MW is supported by NCN Grant No. 2012/05/E/ST2/02352. KR is supported by NCN Grant No. 2015/19/B/ST2/01999. L.K. acknowledges support by the international PhD programme ExQM (Eli	te Network of Bavaria).


\begin{thebibliography}{99}
\bibitem{EPR}  A. Einstein, B. Podolsky, N. Rosen, {\em ``Can Quantum-Mechanical Description of Physical Reality be Considered Complete?'', Physical Review} {\bf 47}, 777 (1935).
\bibitem{Bell}  J. S. Bell, {\em ``On the Einstein-Poldolsky-Rosen paradox'', Physics} {\bf 1}, 195 (1964).
\bibitem{KochenSpecker} S. Kochen and E.P. Specker, {\em ``The problem of hidden variables in quantum mechanics'', Journal of Mathematics and Mechanics}. {\bf 17}, 59 (1967).
\bibitem{Communication} \v{C}. Brukner, M. \.Zukowski, J.-W. Pan, and A. Zeilinger, {\em ``Bell’s Inequalities and Quantum Communication Complexity'', Phys. Rev. Lett.} {\bf 92}, 127901 (2004).
\bibitem{Ekert} A. K. Ekert, {\em ``Quantum cryptography based on Bell’s theorem'', Phys. Rev. Lett.} {\bf 67}, 661 (1991).
\bibitem{Peres} A. Peres, {\em ``Separability Criterion for Density Matrices'', Phys. Rev. Lett.} {\bf 77}, 1413 (1996).
\bibitem{Horod} M. Horodecki, P. Horodecki, and R. Horodecki, {\em ``Separability of Mixed States: Necessary and Sufficient Conditions'', Phys. Lett. A} {\bf 223}, 1 (1996).
\bibitem{JC}  M.-D. Choi, {\em ``Completely positive maps on complex matrices'', Linear Alg.
and Its Appl.} {\bf 10}, 285 (1975); A. Jamio\l kowski, {\em ``Linear transformations which preserve trace and semidefinitness of operators'', Rep. Math. Phys.} {\bf 3}, N4, 275 (1972).
\bibitem{WW} R. F. Werner and M. M. Wolf, {\em ``All-multipartite Bell-correlation inequalities for two dichotomic observables per site'', Phys. Rev. A} {\bf 64}, 032112 (2001).
\bibitem{WZ} H. Weinfurter and M. \.Zukowski, {\em ``Four-photon entanglement from down-conversion'', Phys. Rev. A} {\bf 64} 010102(R)  (2001).
\bibitem{ZB} M. \.Zukowski and \v{C}. Brukner, {\em ``Bell’s Theorem for General N-Qubit States'', Phys. Rev. Lett.} {\bf 88}, 210401 (2002).
\bibitem{Nagata} K. Nagata, W. Laskowski, M. Wie\'sniak, and M. \.Zukowski, {\em ``Rotational Invariance as an Additional Constraint on Local Realism'', Phys. Rev. Lett.} {\bf 93}, 230403 (2004).
\bibitem{WL} P. Badzi\c{a}g, \v{C}. Brukner, W. Laskowski, T. Paterek, and M. \.Zukowski, {\em ``Experimentally Friendly Geometrical Criteria for Entanglement'', Phys. Rev. Lett.} {\bf 100}, 140403  (2008).
\bibitem{WIESNMARU} M. Wie\'sniak and K. Maruyama, {\em ``Package of facts and theorems for efficiently generating entanglement criteria for many qubits'', Phys. Rev. A} {\bf 85}, 062315 (2012).
\bibitem{Lukas} L. Knips, C. Schwemmer, N. Klein, M. Wie\'sniak, and H. Weinfurter, {\em ``Multipartite entanglement detection with minimal effort''}, arXiv:1412.5881.
\bibitem{WLask} R. Weinar, W. Laskowski, M. Paw\l owski, {\em ``Activation of entanglement in teleportation, J. Phys. A} {\bf 46}, 435301 (2013). 
\bibitem{complementarity} G. Jaeger, A. Shimony, and L. Vaidman, {\em ``Two interferometric complementarities'', Phys. Rev. A} {\bf 51}, 54 (1995).
\bibitem{MUBS1} W. K. Wootters and B. D. Fields {\em ``Optimal state-determination by mutually unbiased measurements'', Ann. Phys. (N.Y.)} {\bf 191}, 363 (1989).
\bibitem{MUBS2} I. D. Ivanovi\'c, {\em ``Geometrical description of quantal state determination'', J. Phys. A} {\bf 14}, 3241 (1981).
\bibitem{Multicomponent} J.-L. Chen, Ch.-F. Wu, L. C. Kwek, D. Kaszlikowski, M. \.{Z}ukowski, and C. H. Oh, {\em ``Multicomponent Bell inequality and its violation for continuous-variable systems'', Phys. Rev. A} {\bf 71}, 032107 (2005).
\end{thebibliography}
\end{document}